\documentclass[reqno,11pt]{amsart}
\usepackage{amsmath, latexsym, amsfonts, amssymb, amsthm, amscd}
\usepackage{graphics,epsf,psfrag}
\numberwithin{equation}{section}

\newtheorem{theorem}{Theorem}[section]
\newtheorem{lemma}[theorem]{Lemma}
\newtheorem{proposition}[theorem]{Proposition}

\newtheorem{rem}[theorem]{Remark}

\newcommand{\R}{\mathbb{R}}

\renewcommand{\tilde}{\widetilde}

\newcommand{\cF}{{\ensuremath{\mathcal F}} }
\newcommand{\cG}{{\ensuremath{\mathcal G}} }

\newcommand{\cE}{{\ensuremath{\mathcal E}} }

\newcommand{\cM}{{\ensuremath{\mathcal M}} }

\newcommand{\cD}{{\ensuremath{\mathcal D}} }
\newcommand{\cU}{{\ensuremath{\mathcal U}} }
\newcommand{\cV}{{\ensuremath{\mathcal V}} }

\newcommand{\bE}{{\ensuremath{\mathbf E}} }

\DeclareMathSymbol{\leqslant}{\mathalpha}{AMSa}{"36} 
\DeclareMathSymbol{\geqslant}{\mathalpha}{AMSa}{"3E} 
\DeclareMathSymbol{\eset}{\mathalpha}{AMSb}{"3F}     
\newcommand{\dd}{\,\text{\rm d}}             
\newcommand{\mintwo}[2]{\min_{\substack{#1 \\ #2}}} 

\newcommand{\bbN}{{\ensuremath{\mathbb N}} }

\newcommand{\bbP}{{\ensuremath{\mathbb P}} }

\newcommand{\bbR}{{\ensuremath{\mathbb R}} }
\newcommand{\bbS}{{\ensuremath{\mathbb S}} }

\newcommand{\bbZ}{{\ensuremath{\mathbb Z}} }

\newcommand{\gd}{\delta}
\newcommand{\gep}{\varepsilon}       
\newcommand{\gp}{\varphi}
\newcommand{\gr}{\rho}

\newcommand{\gl}{\lambda}

\newcommand{\gs}{\sigma}

\makeatletter
\def\captionfont@{\footnotesize}
\def\captionheadfont@{\scshape}

\long\def\@makecaption#1#2{%
  \vspace{2mm}
  \setbox\@tempboxa\vbox{\color@setgroup
    \advance\hsize-6pc\noindent
    \captionfont@\captionheadfont@#1\@xp\@ifnotempty\@xp
        {\@cdr#2\@nil}{.\captionfont@\upshape\enspace#2}%
    \unskip\kern-6pc\par
    \global\setbox\@ne\lastbox\color@endgroup}%
  \ifhbox\@ne 
    \setbox\@ne\hbox{\unhbox\@ne\unskip\unskip\unpenalty\unkern}%
  \fi
  \ifdim\wd\@tempboxa=\z@ 
    \setbox\@ne\hbox to\columnwidth{\hss\kern-6pc\box\@ne\hss}%
  \else 
    \setbox\@ne\vbox{\unvbox\@tempboxa\parskip\z@skip
        \noindent\unhbox\@ne\advance\hsize-6pc\par}%
\fi
  \ifnum\@tempcnta<64 
    \addvspace\abovecaptionskip
    \moveright 3pc\box\@ne
  \else 
    \moveright 3pc\box\@ne
    \nobreak
    \vskip\belowcaptionskip
  \fi
\relax
}
\makeatother
\def\writefig#1 #2 #3 {\rlap{\kern #1 truecm
\raise #2 truecm \hbox{#3}}}

\newcommand{\bra}{\left \langle}
\newcommand{\ket}{\right \rangle}
\newcommand{\bbra}{\langle \! \langle}
\newcommand{\kket}{\rangle \! \rangle}

\newcommand{\qhat}{{\hat{q}}}
\newcommand{\im}{{\mathrm i}}

\begin{document}
\title[Kuramoto model and mean field spin XY dynamics]{Dynamical aspects of  mean field plane rotators \\
and  the Kuramoto model}
\date{\today}

\author{Lorenzo Bertini}
\address{	 SAPIENZA Universit\`a di Roma, 
Dipartimento di Matematica
G. Castelnuovo,
P.le Aldo Moro, 2 - 00185 Roma - Italia 
}
\email{bertini\@@mat.uniroma1.it}

\author{Giambattista Giacomin}
\address{
  Universit{\'e} Paris Diderot (Paris 7) and Laboratoire de Probabilit{\'e}s et Mod\`eles Al\'eatoires (CNRS),
U.F.R.                Math\'ematiques, Case 7012 (site Chevaleret)
             75205 Paris
              - France
}
\email{giacomin\@@math.jussieu.fr}
\author{Khashayar Pakdaman}
\address{
  Institut Jacques Monod,
Universit\'e Paris Diderot - CNRS,
B\^at. Buffon - 15 rue H\'el\`ene Brion
75013 Paris - France
}
\email{pakdaman\@@ijm.univ-paris-diderot.fr}

\begin{abstract}
The Kuramoto model has been introduced in order to describe 
synchronization phenomena observed in groups of cells, individuals, 
circuits, etc...   
We look at the Kuramoto model with white noise 
forces: in mathematical terms it is  a set of $N$
oscillators, each driven by an independent Brownian motion with 
a constant drift, that is each oscillator has its own frequency, which, in general, changes from
one oscillator to another (these frequencies  are usually taken to be 
random and they may be viewed as a quenched disorder).
The interactions between oscillators are of long range type (mean field). 
We review some results on the Kuramoto model from
a statistical mechanics standpoint: we give in particular 
necessary and sufficient conditions for  reversibility
and we point out a formal analogy, in the $N \to \infty$ limit, with
local mean field models with conservative dynamics (an analogy that is exploited 
to identify in particular a Lyapunov functional in the reversible set-up).
We then focus on the reversible Kuramoto model with
sinusoidal interactions in the $N \to \infty$ limit and 
analyze the stability of the non-trivial stationary profiles
arising when the interaction parameter $K$ is larger
than its critical value $K_c$. 
We provide an  analysis of the linear operator 
describing the time evolution in a neighborhood 
of the synchronized profile: we exhibit a Hilbert space in which
this operator has a self-adjoint extension and we establish, as our main result,   
a spectral
gap inequality for every $K>K_c$.   
 \\
  \\
  2010 \textit{Mathematics Subject Classification:  82C20, 35P15, 60K35
  }
  \\
  \\
  \textit{Keywords: Synchronization, Kuramoto model, (ir)reversibility, mean field Spin XY model,
  spectral gap inequality }
\end{abstract}

\maketitle 

\section{Introduction}
The Kuramoto model ({\sl e.g.} \cite{cf:review,cf:Strogatz}) is defined by the
set of coupled stochastic differential equations:
\begin{equation}
\label{eq:Kuramoto}
\dd \gp _j^\xi (t) \, =\, \xi_j \dd t -
\frac KN \sum_{i=1}^N \sin \left( \gp^\xi_j(t) -\gp^\xi_i (t)\right) \dd t +
\gs \dd w_j(t)\, ,
\end{equation}
for $j=1,2, \ldots, N$, 
where
\begin{enumerate}
\item $\{ w_j (\cdot)\}_{j=1, \ldots, N}$ is a family of independent and identically
distributed standard Brownian motions. We refer to this source of randomness
as {\sl thermal noise}.
\item $\xi=\{ \xi_j \}_{j=1, \ldots, N}$ is a family of independent identically
distributed random variables. This is another source of noise, and we refer to it
as {\sl disorder}.  
\item $K$ is a real parameter and $\gs\ge 0$. 
\end{enumerate}

\smallskip

We stress from now that we consider the stochastic evolution \eqref{eq:Kuramoto}
once a realization of the disorder variables $\xi$ is chosen, so the disorder
is of {\sl quenched } type. Moreover the law of $\{ w_j (\cdot)\}_{j=1, \ldots, N}$ and of the initial
condition (specified below) does not depend on the values of the disorder variables. 

\smallskip

\begin{rem}
\label{rem:h} \rm
It will be at times interesting to discuss the role
of the sine drift in the model. We will therefore 
refer to a $h$-model when 
$K \sin (\cdot)$ is replaced by a smooth, {\sl i.e.} $C^\infty$,  $2\pi$-periodic function
$h(\cdot)$.
\end{rem}
\smallskip

The  variables $\gp^\xi_j(t)$ are actually angles, 
so we focus on $\gp^\xi_j(t) \text{mod} (2\pi)$,
which is an element of $\bbS := \bbR/(2\pi \bbZ)$.
The existence and uniqueness of a unique (strong) solution 
to the system   \eqref{eq:Kuramoto}, 
when the 
 initial condition 
$\{\gp^\xi_j (0)\}_{j=1, \ldots, N}\in \R^N$ 
and $\{ w_j (\cdot)\}_{j=1, \ldots, N}$ are independent
and $\{\gp^\xi_j (0)\}_{j=1, \ldots, N}$ are square integrable random variables,
is a standard result. In our case we may therefore choose 
$\{\gp^\xi_j (0)\}_{j=1, \ldots, N}$ arbitrarily distributed 
provided that it is concentrated on $[0, 2\pi)^N$. 

The main result of this work is on  the model in which there is no disorder, that
is the case in which the law of $\xi_1$ is degenerate, so that $\xi_j=\xi$
for every $j$,
with $\xi $ is a real constant. 
In this case, with the change of variable
$  \gp_j (t) := \gp^\xi _j(t) - \xi t$  
we have 
\begin{equation}
\label{eq:Kuramoto0}
\dd  \gp _j (t) \, =\, -
\frac KN \sum_{i=1}^N \sin \left(  \gp_j(t) - \gp_i (t)\right) \dd t +
\gs \dd w_j(t)\, .
\end{equation}
Disregarding the disorder 
is actually a major simplification first of all because, if $\gs>0$, 
the system \eqref{eq:Kuramoto0} is {\sl reversible} with
respect to the (Gibbs) probability measure
\begin{equation}
\label{eq:Gibbs}
\mu_{N,K} (\dd \gp)
\,:= \, \frac 1{Z_{N,K}}
\exp\left(-\frac{2K}{\gs^2} H_N (\gp)\right)  \gl_N(\dd \gp)\, ,
\end{equation}
where $\gp \in \bbS^N$, $\gl_N$ is the uniform probability measure
on $\bbS^N$ (that is the $N$-fold product of Lebesgue measures normalized
by $(2 \pi)^N$),
\begin{equation}
\label{eq:spinXY}
H_N (\gp) \, :=\, -\frac 1{2N} \sum_{i=1}^N  \sum_{j=1}^N 
\cos \left( \gp_j -\gp_i \right)\, ,
\end{equation}
and $Z_{N, K}:= \int_{\bbS^N}  \exp(-2K\gs^{-2} H_N (\gp))  \gl_N(\dd \gp)$
is the {\sl partition } function.
In fact, the generator $L_{K,N}$ of the dynamics \eqref{eq:Kuramoto0}
acts on twice differentiable functions $F: \bbS^N \to \R$ as
\begin{equation}
L_{K,N} F(\gp) \, =\, \frac{\gs^2}2\sum_{i=1}^N \frac{\partial^2
F(\gp)}{\partial \gp_i^2}  - K
\sum_{i=1}^N  \frac{\partial H_{N} (\gp)}{\partial \gp_i}
 \frac{\partial F(\gp) }{\partial \gp_i} \, ,
\end{equation}
and one directly verifies the symmetry
$\int F\, L_{K,N}G \dd \mu_{K,N}
=\int G\, L_{K,N}F \dd \mu_{K,N}$ for $F,G \in C^2$, which
 implies that $\mu_{K,N}$ is invariant for the dynamics \cite{cf:Spohn,cf:KL}.

The measure $\mu_{K,N}$ is the Gibbs measure of a classical statistical mechanics model:
the mean field spin XY model with single spin state space 
$\bbS$, {\sl i.e.} mean field plane rotators \cite{cf:SFN}.  
\medskip

\begin{rem}
\rm 
\label{eq:wirr}
It is important to notice that the system \eqref{eq:Kuramoto} is not
reversible unless $\xi_i=0$ for every $i$. Even the case $\xi_i=const. \neq 0$ for every $i$
is not reversible, but, as we have argued, it maps to a reversible system. 
Notice in fact that, unless $\xi \equiv 0$, the  transformation $\gp^\xi_j \mapsto
\gp^\xi_j  -\xi t$ maps to a system with time dependent interactions.
This strongly hints to the absence of reversibility and it
 is indeed the case, but proving such a statement
is more delicate: we address this point in Section~\ref{sec:irreversibility} below. 
The aspect that we want to stress here is 
 the  {\sl  disorder induced non-equilibrium character}
of the full Kuramoto model.  
\end{rem}

\medskip

\subsection{Empirical measure and the large $N$ limit.}
Since we focus on the $\gs>0$ case,
there is  no loss in generality in choosing $\gs=1$ and
we will do so from now on.
We introduce the empirical measure
\begin{equation}
\label{eq:emp}
\nu_{N, t}(\dd \theta) \, :=\, \frac 1N 
\sum_{j=1}^N \gd_{\gp_j(t)}(\dd \theta),  
\end{equation}
and observe that, 
by It\^o's rule, for every $F\in C^2 (\bbS)$ and $t>0$
\begin{multline}
\int_{\bbS} F(\theta) \nu_{N, t}(\dd \theta) -
 \int_{\bbS} F(\theta) \nu_{N, 0}(\dd \theta)\, =
 \\
 -K \int_{0}^t
 \int_{\bbS ^2}F^\prime (\theta) \sin (\theta -\theta^\prime)
 \nu_{N, s}(\dd \theta) \nu_{N, s}(\dd \theta^\prime) \dd s +
 \frac 12 \int_{0}^t
 \int_{\bbS }F^{\prime\prime} (\theta)   \nu_{N, s}(\dd \theta) \dd s +
 M_{N, F} (t)\, ,
\end{multline}
where $M_{N, F} (\cdot)$ is a continuous martingale 
with quadratic variation at time $t$, $\langle M_{N,F}\rangle(t)$, 
equal to $N^{-1} \int_0^t \int_{\bbS} 
(F^\prime(\theta))^2 \nu_{N, s} (\dd \theta)\dd s$. Therefore, by Doob's inequality,
for every $T>0$ we have that 
$\bE[\sup_{t \in [0, T] } ( M_{N, F} (t))^2]$  is bounded by 
$\langle M_{N,F}\rangle(T)\le T \Vert F^\prime \Vert_\infty^2 /N$:
this guarantees that the thermal noise disappears as $N \to \infty$,
so that the limit of the empirical measure, if it exists, is not random.
To make this precise
we introduce the space $C^0 ([0,T]; \cM_1(\bbS))$,
where $\cM_1(\bbS)$ are the probability measures on $\bbS$ equipped with the 
topology of the weak convergence, and observe that
if  a subsequence of $\{\nu_{N, \cdot}\}_{N \in \bbN}$ 
(of elements of $C^0 ([0,T]; \cM_1(\bbS))$) converges to a limit 
$\nu_\cdot$,  we have that for $t \in (0, T]$ and every $F\in C^2 (\bbS)$
\begin{multline}
\label{eq:limitweak}
\int_{\bbS} F(\theta) \nu_{ t}(\dd \theta) -
 \int_{\bbS} F(\theta) \nu_{0}(\dd \theta)\, =
   \\
   \frac 12 \int_{0}^t
 \int_{\bbS }F^{\prime\prime} (\theta)   \nu_{s}(\dd \theta) \dd s
 -K \int_{0}^t
 \int_{\bbS ^2}F^\prime (\theta) \sin (\theta -\theta^\prime)
 \nu_{s}(\dd \theta) \nu_{s}(\dd \theta^\prime) \dd s \, .
\end{multline}
This is a weak form of the equation
\begin{equation}
\label{eq:classical}
\partial_t q_t(\theta) \, =\, \frac 12 \frac{ \partial ^2 q_t (\theta)}{\partial \theta^2}
+ K  \frac{\partial}{\partial \theta}\left[\left(
\int_{\bbS} \sin(\theta- \theta^\prime ) q_t(\theta^\prime) \dd \theta^\prime
\right)
q_t(\theta)\right] ,
\end{equation}
when $\nu_t(\dd \theta)= q_t(\theta) \dd \theta$. So that
if we assume that $\nu_{N,0}$ converges to a non random limit and 
 if there is a unique solution to \eqref{eq:limitweak}, then the evolution
 is non random and determined by  \eqref{eq:limitweak}.

More precisely, we have the following:

\medskip
\begin{proposition}
\label{th:Nlim}
If there exists $\nu_0 \in \cM_1(\bbS)$ such that for every $\gep>0$
and every $F \in C^0(\bbS ;  \R)$ we have
\begin{equation}
\lim_{N \to \infty}
\bbP\left( \left\vert \int _{\bbS} F(\theta) \nu_{N,0}(\dd \theta) -
 \int _{\bbS} F(\theta) \nu_{0}(\dd \theta)
 \right\vert > \gep\right) \,=\, 0\, ,
\end{equation}
then for every $t>0$ we have that for every $\gep$ and $F$
\begin{equation}
\lim_{N \to \infty}
\bbP\left( \left\vert \int _{\bbS} F(\theta) \nu_{N,t}(\dd \theta) -
 \int _{\bbS} F(\theta) \nu_{t}(\dd \theta)
 \right\vert > \gep\right) \,=\, 0\, ,
\end{equation}
where $\nu_\cdot$ is the unique solution of \eqref{eq:limitweak}.
Moreover,  for every $t>0$ the measure
$\nu_t$ is absolutely continuous with respect to the Lebesgue measure with (strictly) positive
density $q_t(\cdot )$ and the function 
$(t, \theta) \mapsto q_t(\theta)$, from $(0, \infty ) \times 
\bbS$ to $(0, \infty)$, is smooth and solves 
 \eqref{eq:classical}. 
\end{proposition}

\medskip

Proposition \ref{th:Nlim} is a particular (and particularly easy) case of far more general results 
(see for example \cite{cf:Gartner,cf:Oelsch}). 
The derivation goes along proving tightness of $\{ \nu_{N, \cdot}\}_{N\in \bbN}$
and then proving uniqueness for the limiting equation \eqref{eq:limitweak}.
In our case such an equation is particularly nice and the evolution is smoothing,
so that even if the initial datum is not a function ({\sl i.e} $\nu_0$ is not absolutely
continuous with respect to the Lebesgue measure) or it is not smooth, 
$\nu_t(\dd\theta)= q_t(\theta )\dd \theta$ and $q_t(\cdot) \in C^\infty(\bbS)$
for every $t>0$.
These analytic aspects are taken up with more details in
Section~\ref{sec:nonlinear} (Proposition~\ref{th:nonlinear})
   
\medskip

\begin{rem}\rm
\label{rem:withdisorder}
It is however important to recall here that  Proposition~\ref{th:Nlim} can be generalized
to cover the disordered case \eqref{eq:Kuramoto}. We refer to
\cite{cf:dPdH,cf:dH} for precise statements, but, roughly, if the law of the random variable $\xi_1$
is denoted by $\mu (\dd \xi)$ (let us for example assume
that $\xi_1$ is bounded), the empirical average at time $t>0$
converges as $N \to \infty$ to a measure with density
 $\int_\R q_t(\theta; \xi) \mu(\dd\xi)$, where $\{q_t(\theta; \xi)\}_{t\ge 0, \theta \in \bbS,
 \xi \in \R}$ 
is the unique solution to 
\begin{equation}
\begin{split}
\label{eq:dPdH}
\partial_t q_t (\theta; \xi ) \, &=\, \frac 12 \frac{\partial^2 q_t (\theta; \xi)}{\partial \theta^2}
+ \frac{\partial}{\partial \theta} \left[ \left(\int_\R \left(
\int_{\bbS} K \sin(\theta- \theta^\prime ) q_t(\theta^\prime; \xi^\prime) \dd \theta^\prime\right) \mu(\dd \xi^\prime) + \xi
\right)
q_t(\theta; \xi)\right],
\\
q_0(\theta; \xi) \, &=\, \frac{\dd \nu_0(\dd \theta)}{\dd\theta}\, , 
\end{split}
\end{equation}
for every $\xi$ in the support of $\mu$. We have of course  assumed, for simplicity, that $\nu_0$ is absolutely
continuous with respect to the Lebesgue measure.
\end{rem}

\medskip

\begin{rem}\rm
\label{rem:Kac}
The Kuramoto limit evolution \eqref{eq:classical} comes up also 
as {\sl mesoscopic scaling limit} for the density of Ka\v{c} models 
with conservative dynamics
\cite{cf:BL,cf:GL},
when the interaction potential is chosen equal to $\cos(\cdot)$.
This is just a formal analogy, but it is going to be crucial for the sequel. 
\end{rem}
 
\subsection{Stationary profiles}
By the regularizing character of the evolution (Proposition~\ref{th:nonlinear}), 
the stationary solutions 
to \eqref{eq:limitweak} coincide with the stationary solutions 
to \eqref{eq:classical} (we are of course interested only in non-negative solutions of 
total mass equal to one:  Proposition~\ref{th:nonlinear} guarantees also
the positivity of stationary solutions). Let us notice moreover that if $\qhat(\cdot)$ is a stationary
solution, then $\qhat  (\cdot+\theta_0)$ is a stationary solution too, for arbitrary choice of $\theta_0$. This is  due to the invariance of \eqref{eq:Kuramoto0} under
rotations (that can of course be read also out of \eqref{eq:spinXY}).
Note that $\qhat  (\cdot)= 1/(2\pi)$ is a solution to \eqref{eq:Kuramoto0}, regardless of the
value of $K$ but there may be more solutions: in fact every stationary solution
can be written as  
 $\qhat  (\cdot+\theta_0)$ for some $\theta_0\in [0,2\pi)$
and 
\begin{equation}
\label{eq:stat}
\qhat  (\theta) \, :=\,  \frac{\exp(2Kr \cos(\theta)}{\int_{\bbS} \exp(2Kr \cos(\theta^\prime) \dd \theta^\prime},
\end{equation} 
 with $r$ a non-negative solution to
 \begin{equation}
 \label{eq:fixr}
 r\, := \, \Psi \left( 2Kr \right), \  \ 
 \   
 \text{ with } \ \ \Psi(x)\, :=\, \frac{\int_{\bbS} \cos(\theta) \exp(x \cos (\theta))\dd \theta}{\int_{\bbS}  \exp(x \cos (\theta))\dd \theta}.
 \end{equation}
In general, there is more than one solution to \eqref{eq:fixr}: in fact, there
can be at most two, more precisely there is 
only the trivial solution $r=0$ for $K\le 1$ and 
there is also a second solution $r>0$ if $K>1$. This is because
$\Psi ^\prime (0)=1$ and because 
$\Psi(\cdot):[0, \infty)\to [0, 1)$ is strictly concave \cite{cf:Pearce}.
In terms of stationary solutions, this means that
for $K\le 1$ only the flat ({\sl incoherent}) profile $1/(2\pi)$
is stationary, while for $K>1$ also $\{\qhat (\cdot + \theta_0)\}_{\theta_0 \in \bbS}$
is a family of stationary solutions (they are the  solutions 
that exhibit the {\sl coherence} or {\sl synchronization} of the system).

The result we just stated, that is \eqref{eq:stat}-\eqref{eq:fixr}, is a classical one in the sense
that it is of course closely linked to the solution of the mean field
planar rotator model \cite{cf:SFN} (result completed by 
the concavity result proven in \cite{cf:Pearce}). It is however worthwhile
recalling the proof: every stationary solution $\qhat$ of \eqref{eq:classical} satisfies
\begin{equation}
\label{eq:eqst1}
\frac 12 \qhat^{\prime}(\theta)
+ K \left(
\int_{\bbS} \sin(\theta- \theta^\prime ) \qhat(\theta^\prime) \dd \theta^\prime
\right)
\qhat(\theta)\, =\, C\, ,
\end{equation}
for some constant $C$. 
Since we know that $\qhat(\cdot)>0$, then \eqref{eq:eqst1} yields
\begin{equation}
\label{eq:eqst2}
\frac 12 \left( \log \qhat (\theta) \right) ^\prime- 
K \left(
\int_{\bbS} \cos(\theta- \theta^\prime ) \qhat(\theta^\prime) \dd \theta^\prime
\right)^\prime \, =\, \frac{C}{\qhat (\theta)}\, ,
\end{equation}
which implies $C=0$. At this point, by playing on the
rotation symmetry,  we may assume that 
$\int_\bbS \qhat (\theta) \sin(\theta)\dd \theta=0$, so that any stationary 
non-negative solution $\qhat (\cdot)$ 
with prescribed first Fourier cosine coefficient $\int_{\bbS} \qhat (\theta) \cos (\theta)\dd \theta$
equal to $r$
satisfies 
\begin{equation}
\label{eq:eqst3}
\frac 12  \qhat (\theta) ^\prime- 
K r \cos(\theta) \qhat (\theta)\,  =\,0 .
\end{equation}
A solution to \eqref{eq:eqst3} is proportional to 
$\exp(2Kr \cos (\theta)$. By normalizing ($\qhat (\cdot)$ is a probability density)
and recalling the constraint 
on the first Fourier cosine coefficient we get to
\eqref{eq:stat}--\eqref{eq:fixr}.

\medskip

\begin{rem}
\label{rem:Sakaguchi}
\rm
Remarkably, a generalization of  \eqref{eq:stat}--\eqref{eq:fixr}
holds also in the disordered case \cite{cf:Sakaguchi}. 
The key to such a derivation, like in the step above, is in the identification
of the order parameter $r$, that captures the degree of {\sl coherence}
(or {\sl synchronization}) of the oscillators. In statistical mechanics terms 
this is nothing but the fact that the Hamiltonian $H_N(\gp)$ may be rewritten
as $(1/2N)\sum_{i,j} S_i \cdot S_j=(1/2N)(\sum_{i} S_i )^2$, with 
$S_i= (\Re \exp( \im \gp_i),\Im \exp( \im \gp_i))$.
However, 
if one considers an $h$-model ({\sl cf.} Remark~\ref{rem:h}),
 the Hamiltonian cannot be expressed any longer
as a function of the total magnetization $\sum_i S_i$.
For the identification of the order parameter in this more general
context we refer to \cite{cf:Daido}. 
\end{rem}

\subsection{The gradient flow viewpoint} 
For our purposes the following fact is
of crucial importance: 
\eqref{eq:classical} can be rewritten in the {\sl gradient form}
\begin{equation}
\label{eq:classical2}
\partial_t q_t(\theta) \, =\, \nabla\left[
q_t(\theta) 
\nabla\left(
\frac{\gd \cF(q_t)}{\gd q_t(\theta)}\right)
\right] \, , 
\end{equation}
 where we use $\nabla $ for $\partial _\theta$ 
 for visual impact, $\gd \cG(q)/ \gd q(\theta)$ is the standard $L^2$ Fr\'echet derivative
 of the functional $\cG$ and 
 \begin{equation}
 \label{eq:F}
\cF(q)\, :=\, \frac 12 \int _{\bbS} 
q(\theta) \log q(\theta) \dd \theta
- \frac K2 \int_{\bbS^2} \cos (\theta- \theta^\prime) q(\theta) 
q(\theta^\prime) \dd \theta \dd \theta^\prime.
\end{equation}
Note that $\cF: L^2(\bbS) \to \R$ is Fr\'echet differentiable at
$q(\cdot)$ for example if $q(\cdot)$ is continuous and 
$q(\cdot)>0$ and Proposition~\ref{th:nonlinear} guarantees 
that the evolution may be cast in the form \eqref{eq:classical2}
for $t>0$.
A direct consequence of \eqref{eq:classical2} is that
\begin{equation}
\label{eq:Lyap}
\frac{\partial \cF(q_t)}{\partial t} \, =\, - \int_{\bbS} 
q_t(\theta)  \left( \nabla\frac{\gd \cF(q_t)}{\gd q_t(\theta)}\right)^2 \dd \theta \, \le \, 0
\, .
\end{equation}
A first consequence of this observation is:
\medskip

\begin{proposition}
\label{th:rotating}
If there exists $t_2>t_1\ge 0$ such that $q_{t_1}(\cdot)=q_{t_2}(\cdot)$, then 
there exists a constant $C$ and a value $r$  satisfying \eqref{eq:fixr} such that
$q_t(\cdot)=\hat q(\cdot +c)$ for every $t\ge t_1$.
\end{proposition}
\medskip

\noindent
{\it Proof.}
By Proposition~\ref{th:nonlinear} (guaranteeing smoothness and positivity of
the solution) and by \eqref{eq:Lyap} we see that $\cF(q_t)$ is constant for 
$t \in [t_1,t_2]$, so that
$\nabla\gd \cF (q_t) /\gd (q_t (\theta))=0$ for every $\theta$ and $t \in [t_1,t_2]$.
But this implies that
\begin{equation}
\label{eq:eqst2.1}
\frac 12 \nabla \left( \log q_t (\theta) \right) 
+ 
K \left(
\int_{\bbS} \sin(\theta- \theta^\prime ) q_t(\theta^\prime) \dd \theta^\prime
\right) \, =\, 0\, ,
\end{equation}
which is precisely \eqref{eq:eqst2} with $C=0$.
Therefore for every $t \in [t_1,t_2]$ there exists  a constant $\gamma (t)$ such that
$q_t(\theta)=\qhat ( \theta+ \gamma(t))$, with
$\qhat(\cdot)$ as in \eqref{eq:stat}-\eqref{eq:fixr}. 
Since $\qhat ( \cdot+ \gamma(t_1))$ is a stationary solution,
the claim follows. 
\qed

\medskip

Two observations are in order:
\smallskip

\begin{enumerate}
\item
\label{above}
Proposition~\ref{th:rotating} generalizes to the non-disordered
$h$-model, when the latter is reversible (see Section~\ref{sec:irreversibility}),
in the sense that the hypotheses imply $\nabla (\gd \cF/ \gd q_t )=0$
and this condition identifies all the stationary solutions.
\item
Proposition~\ref{th:rotating} 
shows that there is no non-trivial stationary solution to
\eqref{eq:Kuramoto} when $\xi_j \equiv \xi $, $\xi$ a non-zero constant. 
This is simply because, otherwise, we would have
a solution to \eqref{eq:Kuramoto0} of the form
$q(\cdot - t \xi)$, with $q(\cdot) $ non-constant, which violates Proposition~\ref{th:rotating}. 
This 
is of interest also because it is not clear that Proposition~\ref{th:rotating} 
generalizes to disordered models. 
Clarifying the link between 
non-reversibility and coexistence of stationary and 
{\sl rotating} solutions appears also to be an intriguing
question. 
\end{enumerate}

\subsection{On synchronization stability}
The main result that we present addresses the important issue of the stability
of  the non-trivial stationary profiles $\qhat(\cdot)$, more precisely of
the stability of the invariant {\sl manifold}
$\{ \qhat (\cdot+\theta_0)\}_{\theta_0\in \bbS}$.
In the literature we find a full analysis of incoherence stability \cite{cf:StMi}
(also in presence of disorder)
as well as an analysis of synchronized profiles as bifurcation
from the incoherent $1/2\pi$ profile (we refer to \cite{cf:review} and the several references 
therein). Our aim is to have a detailed non-perturbative analysis of the
linearized evolution operator in the non disordered case, for every $K>K_c=1$.

To address such an issue we observe that
the linearized evolution $u_t (\cdot)$ around 
$\qhat (\cdot)$ obeys the equation $\partial_t u_t (\theta)=
 L_{\qhat} u_t(\theta) $ with $L_{\qhat}$ a linear operator
 with domain 
 $D( L_{\qhat} ):= \{ 
 C^2( \bbS; \R ):\, 
 \int_{\bbS} u(\theta) \dd \theta
 =0\}
 $
 defined as
 \begin{equation}
 \label{eq:operator}
 L_{\qhat} u (\theta)\, =\, 
 \frac12 \Delta u(\theta) + K \nabla \left[
 \qhat (\theta) 
 \int_{\bbS} \sin \left( \theta- \theta^\prime\right) u \left(\theta^\prime\right)
 \dd \theta^\prime +
 u(\theta) 
 \int_{\bbS} \sin \left( \theta- \theta^\prime\right) \qhat \left(\theta^\prime\right)
 \dd \theta^\prime
 \right] .
 \end{equation}
It is easy to verify that
$L_{\qhat} \qhat ^\prime =0$, and this corresponds 
to the rotation invariance of the problem. However, what we are going to prove is that the 
remaining part of the spectrum is also real and it lies 
on the negative semi-axis. In order to make precise statements
about $L_{\qhat}$ we  introduce the Hilbert space
$H_{-1,1/\qhat}$ of distributions $u$ such that
$u= \cU ^{\prime}$, with $\cU \in L^2 (\bbS; \R)$.
Of course the derivative is taken in the sense of distributions
and $\cU$ is determined, given $u$, only up to
a constant: we remove this uncertainty by stipulating
that $\int_{\bbS} (\cU (\theta)/ \qhat(\theta))\dd \theta =0$.
The norm of $u \in H_{-1,1/\qhat}$ is defined by 
\begin{equation}
\label{eq:H-1}
\Vert u \Vert_{-1, 1/\qhat} ^2\, :=\, \int_{\bbS} \frac{\cU (\theta)^2}{\qhat(\theta)}\dd \theta , 
\end{equation}
and the scalar product of $u$ and $v\in H_{-1,1/\qhat}$ is going to be denoted
by $\bbra u, v \kket$: it is of course equal to $ \int_{\bbS}
(\cU(\theta)\cV(\theta)/\qhat (\theta)) \dd \theta$, with  definition of $\cV$ in analogy
with $\cU$.
We will come back in the next section  with more on $H_{-1,1/\qhat}$,
but what one can verify directly is that $D( L_{\qhat} )$ and $L_{\qhat} D( L_{\qhat} )$  
are subsets of $ H_{-1,1/\qhat}$
and that $L_{\qhat}$ is symmetric as an operator on   $H_{-1,1/\qhat}$, that is
\begin{equation}
\bbra u, L_{\qhat} v \kket \, =\, \bbra v, L_{\qhat} u \kket ,
\end{equation}
for every $u,v \in D( L_{\qhat} )$ (for an explicit expression see \eqref{eq:sym}). 
We will actually prove (Proposition~\ref{th:selfadj}) that $L_{\qhat}$ is essentially self-adjoint. Moreover:

\medskip

\begin{theorem}
\label{th:spectrum}
The spectrum of (the self-adjoint extension of) $L_{\qhat}$ is pure point and it lies in 
$(-\infty, 0]$. The value $0$ is in the spectrum, with one-dimensional eigenspace 
(spanned, as we have seen, by $\qhat^\prime$) and the distance between zero
and the rest of the spectrum is of at least
\begin{equation}
\label{eq:spectralgapth}
\gl(K)\, :=\, 
 \frac{ \left(1-K(1-r^2)\right) \left(1- (I_0(2Kr))^{-2}\right)}{2K r^2 \exp(8Kr)+ \exp(4Kr) \left(1- (I_0(2Kr))^{-2}\right)}\, >\, 0.
\end{equation}
\end{theorem}

\medskip

 We stress that Theorem~\ref{th:spectrum}
holds as soon as there is a non-trivial solution $r$ to \eqref{eq:fixr}, that is for every $K>1$. We have:
\begin{equation}
\gl(K)\stackrel{K \searrow 1} \sim \frac{K-1}2 \ \ \  \text{ and  } \ \ \ 
\gl(K)\stackrel{K \to  \infty} \sim \frac{\exp(-8K+2)}{4K}.
\end{equation}
Numerically increases till 
$K=1.033\ldots$, where it reaches 
the value $0.0028\ldots$,
and then it decreases.

\medskip

The paper is organized as follows: 
in Section~\ref{sec:synch} we study $L_{\hat q}$ and  prove 
a spectral gap inequality, the essential self-adjontness of the operator
and the fact that the spectrum is pure point.
The nonlinear evolution properties mentioned in this introduction 
are treated in Section~\ref{sec:nonlinear} and 
the (ir)reversibility  issues are considered in Section~\ref{sec:irreversibility}.

\section{Synchronization stability}
\label{sec:synch}

In this section we prove the main result (Theorem~\ref{th:spectrum}).
We assume $K>1$ and, 
for simplicity, we  drop the hat from $\qhat(\cdot)$, so that a stationary
solution is denoted by 
 $q(\cdot)$.
 
\subsection{Some properties of the stationary profile}
\begin{figure}[htp]
\begin{center}
\leavevmode
\epsfxsize =11 cm
\psfragscanon
\psfrag{v}[c]{$K$}
\psfrag{rK}[c]{$r(K)$}
\epsfbox{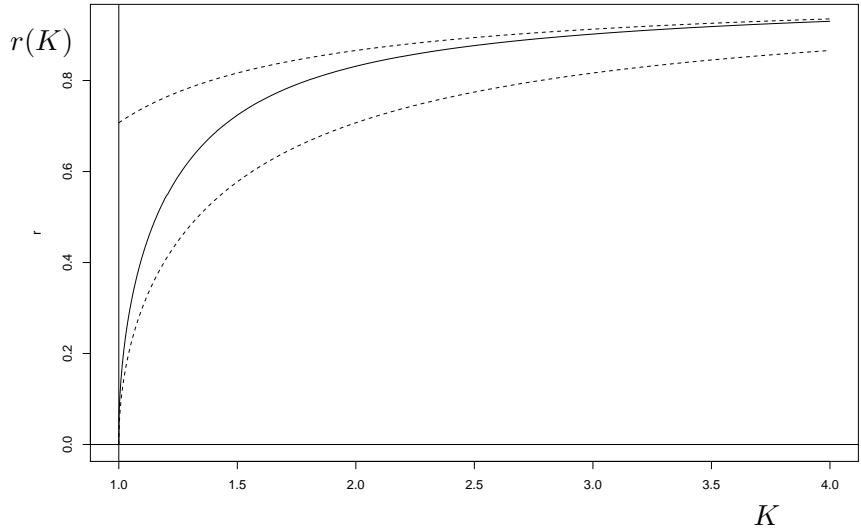}
\end{center}
\caption{\label{fig:Bessel} 
The function $K \mapsto r(K)$ and (dashed lines) the bounds of \eqref{eq:Bessel2}. The upper
bound is proven, the lower bound is verified for $K$ close to 1 and $K$ large.}
\end{figure}

 We first rewrite 
 \eqref{eq:stat}-\eqref{eq:fixr}
by using the Bessel function notation:
\begin{equation}
\label{eq:q}
q(\theta)\, :=\, \frac1 {2\pi I_0(2Kr)} \exp\left( 2Kr \cos(\theta)\right),
\end{equation}
and  $r\in (0,1)$ is the unique 
positive solution of
\begin{equation}
\label{eq:r}
r\,:=\, \Psi (2Kr), \ \  \text{ with } \ \Psi(x)\, :=\, \frac{I_1(x)}{I_0(x)}.
\end{equation}
We have used the standard notation for the modified Bessel functions
of order $0$ and $1$,
explicitly
\begin{equation}
I_\nu (x) \, :=\, \frac 1{2\pi}\int_{0}^{2\pi} (\cos (\theta))^\nu\exp\left( x \cos(\theta)\right) \dd \theta, \ \ \ \ \  \ \ \text{ for } \nu =0,1.
\end{equation}
As already mentioned before, uniqueness of $r$ is a non-trivial fact 
that follows from  \cite[Lemma~4]{cf:Pearce}, that establishes in particular
the concavity of $\Psi(\cdot)$ on the positive semi-axis. 
One can therefore define, via \eqref{eq:r}), the function $[1, \infty)\ni K \mapsto r(K) \in [0,1)$
(one sets $r(1):=0$ by continuity).
We have that,
 for $K>1$, $1-K(1-r^2) \in (0,1/2)$ or (equivalently)
\begin{equation}
\label{eq:Bessel2}
\sqrt{1 -\frac 1K} \, < r(K) \, <\,  \sqrt{1 -\frac 1{2K}} .
\end{equation}
These bounds are easily checked for $K$ close to $1$
and $K$ large, and and the numerical plots of the
three functions appearing in \eqref{eq:Bessel2}
ca be found in Figure~\ref{fig:Bessel}. 
We could not find quick proofs of \eqref{eq:Bessel2}:
 a proof of 
the upper bound is a byproduct of one of the arguments that we develop below (see the 
 proof of Lemma~\ref{th:prestable}), while we prove here
the lower bound by using Bessel functions properties. 

\smallskip

\noindent
{\it Proof of \eqref{eq:Bessel2}, lower bound.}
By the change of variables $(r,K) \mapsto (r, 2Kr)=: (r,y)$
we see that what we have to prove
 is equivalent to showing that
\begin{equation}
\label{eq:tbsPsi2}
\Psi^2(y) + \frac{2}y \Psi(y) -1 \stackrel{y>0}>  0\,  ,
\end{equation}
Apply now the identity \cite{cf:Watson}
\begin{equation}
\frac{I_1(y)}{I_{0}(y)}\, =\, \frac{y}{2} \left( \frac 1{
1+  \frac{y}{2} \frac{I_{2}(y)}{I_{1}(y)}
} \right), 
\end{equation}
so that
\begin{equation}
1-\frac 2y \Psi(y)\, =\, \frac 1{1+ \frac 2y\frac{I_1(y)}{I_2(y)}}\, (>0), 
\end{equation}
and therefore
\eqref{eq:tbsPsi2}  is equivalent to
\begin{equation}
\Psi^2(y) \left( 1+ \frac 2y \frac{I_1(y)}{I_2(y)}\right) \, 
=\,
\frac{I_1(y)^2}{I_0(y)I_2(y)}\, >\,1\, , 
\end{equation}
where the intermediate step follows from the identity 
$yI_2(y)+2I_1(y) -yI_0(y)=0$ \cite{cf:Watson}.
But this is equivalent to
$I_1(y)/I_0(y)>I_2(y)/I_1(y)$ for $y>0$, a fact that is proven in 
\cite[(3.8)]{cf:JoBi}.
 \qed

\medskip

In the sequel we will also use the notations 
\begin{equation}
\label{eq:J}
J(\theta)\,:=\,  -K \sin (\theta) \ \ \ \text{ and  }
\ \ \ \tilde J(\theta):= K \cos(\theta)\, ,
\end{equation}
so that
\begin{equation}
\label{eq:convstat}
\frac{q^\prime}{2q} \, =\, J*q\, ,
\end{equation}
and with this change of notation \eqref{eq:operator} reads
\begin{equation}
\label{eq:operator2}
L_q u \, =\, 
\frac 12 u ^{\prime\prime} - \left(
q \left(J*u\right) + u \left(J*q\right)
\right)^\prime.
\end{equation}

\subsection{Rigged Hilbert spaces and  $H_{-1,1/q}$}

We now introduce a  
{\sl rigged Hilbert spaces} structure 
\cite[pp.~81-82]{cf:Brezis}. The {\sl pivot} (Hilbert) space is
$H:=\{ u \in L^2(\bbS): \, \int u =0\}$ (of course the scalar product is $(u,v):= \int u v$
and the norm is denoted by $\Vert\cdot \Vert_2$).
The second Hilbert space we consider is $V:= H_{1,q}$, closure of the set of periodic
$C^1$ functions $u$ such that $\int u=0$
with respect to the norm
\begin{equation}
\Vert u \Vert_{1,q}\, :=\, \sqrt{\int_{\bbS} (u')^2 q}, 
\end{equation}
so that $V\subset H$ and
the canonical injection of $V$ into $H$ is continuous (by the Poincar\'e inequality).
Note that $V$ is dense in $H$.
We consider then the dual space $V'$ of $V$
and the duality functional in $V'$ defined by 
$\gp _u(v):= (u,v)$ for every given $u \in H$ ($\gp_u: V \to \R$).
We can define $T: H \longrightarrow V'$ by setting
$Tu:= \gp_u$. One can then show that $T(H)$ is dense in $V'$ and 
$T$ injects  $H$ into $V'$ in a continuous way  \cite[p.~82]{cf:Brezis}.
This injection allows considering $H$ as a subset of $V'$, by identifying $u$
and $Tu$. Moreover if $u\in H$ we have
that  $\Vert u \Vert_{V'}=\Vert Tu \Vert_{V^\prime}$ can be made 
more explicit:
given $u\in H$ we call $\cU$ the primitive of $H$ such that
$\int U/q=0$  and we observe that
\begin{equation}
\Vert u \Vert _{V'}\, =\, \sup_{v \in V} \frac{(u,v)}{\Vert v \Vert_V}\, =\, 
\sup_{v \in H_{1,q}} \frac{\int \cU v'}{\sqrt{\int q(v')^2} } \, = \, \sqrt{\int \frac{\cU^2}q},
\end{equation}
where the last step follows on one hand by  the Cauchy-Schwarz inequality 
(establishing the upper bound) and 
by choosing $v'= \cU/q$ in the supremum (establishing the lower bound). 

As already mentioned in the introduction, the scalar product in 
  $V'=H_{-1,1/q}$ is denoted by $\bbra\cdot, \cdot \kket$. 
 We observe also that these steps allow the precise identification of the functions in 
  $H_{-1, 1/q}$: $u \in H_{-1, 1/q}$ if and only if $u=U'$ (in the sense of distributions),
  with $ U \in L^2(\bbS)$.

 \medskip  
 
 At this point it is crucial to observe that
 $D(L_q)$ (recall that $D(L_q)$ is  the subset of periodic $C^2$ functions $u$ such that
 $\int u=0$) is dense in $H_{-1, 1/q}$ and that
 for $u,v \in D(L_q)\subset H_{-1, 1/q}$,
 we have (use \eqref{eq:convstat})
\begin{equation}
\label{eq:sym}
\bbra v, L_q u \kket\, =\,  \bbra L_q v, u \kket \, =\, -\frac 12
\int_{\bbS} \frac {uv}q + \left( v, \tilde J*u \right).
\end{equation}  
In words: $L_q$ is a symmetric operator on $H_{-1, 1/q}$.

\subsection{Estimates on the Dirichlet form}

It is going to be useful to introduce also the Hilbert space $L^2_{1/q}$,
which coincides with $L^2(\bbS)$ as a set of functions, but we equip it
with the scalar product
 \begin{equation}
\bra u, u\ket \, := \,(u, u/q).
\end{equation}
We therefore introduce, for $u \in D(L_q)$, the Dirichlet form
$\cD(u):=  - \bbra u, L_q u \kket$. By \eqref{eq:sym} we have 
\begin{equation}
\label{eq:cD}
\cD (u)\, =\, \frac 12 \bra u, u\ket -  \left(
u , \tilde J * u \right).
\end{equation}
\medskip

Our aim is to bound from below $\cD(u)$ and we start with two technical lemmas.
The first one yields the spectral decomposition of 
$(u, \tilde J*u)$, viewed as a quadratic form on $L^2_{1/q}$.

\medskip

\begin{lemma}
\label{th:decomp}
We have the orthogonal decomposition 
\begin{equation}
\label{eq:decomp}
L^2_{1/q}\, =\, V_0 \oplus^{ \bot} V_{1/2} \oplus^{\bot} V_{K-1/2},
\end{equation}
where
\begin{equation}
V_0 \, :=\, \left\{ \theta \mapsto a_0 +\sum_{j \ge 2} (a_j \cos(j\theta)+b_j \sin (j\theta)):
\, \sum_j a_j^2+b_j^2 < \infty\right\}, 
\end{equation}
and both $V_{1/2} $ and $V_{K-1/2}$
are one dimensional subspaces generated respectively
by $\theta \mapsto  \sin(\theta) q(\theta)\, (= -q^\prime(\theta)/2Kr)$
and by  $\theta \mapsto  \cos(\theta) q(\theta)$.
Moreover, when $u \in V_\gl$ we have
\begin{equation}
\label{eq:dcmp2}
 \tilde  J * u \,= \, \frac{\gl}q \, u. 
\end{equation}
\end{lemma}

\medskip

\noindent
{\it Proof.}
The $L^2_{1/q}$-orthogonality statements $V_0 \bot V_{1/2}$ and 
$V_0\bot V_{K-1/2}$ follow directly from the orthogonality
in $L^2$ of the family $\{\cos(j\theta), \sin (j\theta)\}_{j=0,1, \ldots}$.
Instead $V_{1/2}\bot V_{K-1/2}$ because
$\int_{-\pi}^{\pi}q(\theta) \cos(\theta) \sin(\theta)\dd \theta =0$.

The validity of \eqref{eq:dcmp2} follows by direct computation:
for $u\in V_0$
\begin{equation}
\tilde  J * u (\theta) \, =\, K
\cos(\theta)\int_0^{2\pi}  \cos(\theta^\prime)u(\theta^\prime) \dd \theta^\prime
 +K \sin(\theta) \int_0^{2\pi} \sin(\theta^\prime))
u(\theta^\prime) \dd \theta^\prime,
\end{equation}
which is equal to zero because $u$ does not contain the first harmonics.
The other two cases follow by using the same trigonometric identity
and the following two (clearly equivalent) identities:
\begin{equation}
\label{eq:sincos2}
\int_0^{2\pi} q(\theta) \sin ^2 (\theta)\dd \theta \, =\, \frac 1{2K}, \ \ \ \ \ \ 
\int_0^{2\pi} q(\theta) \cos ^2 (\theta)\dd \theta \, =\, \left(1-\frac 1{2K}\right).
\end{equation}
\qed

\medskip

The second lemma is more technical and its interest will become clear in
the proof of Proposition~\ref{th:stable}.

\medskip
\begin{lemma}
\label{th:prestable}
We have
\begin{equation}
\label{eq:claim1}
\min_{u \in V_0: \, \int u =0}
 \bra 1 +{ u},  1 +{ u} \ket
 \, =\,
  \bra 1 +{\hat u},  1 +{\hat u} \ket
 \, 
 =\, (2\pi)^2 \frac{2K-1}{2K(1-r^2)-1}\, ,
\end{equation}
where
\begin{equation}
\hat u (\theta) \, =\, -1 -  \frac{2\pi \left(1-\frac1{2K}\right)}{\left(
r^2- \left(1-\frac1{2K}\right) 
\right)} \, q(\theta)\, + 
\frac{2\pi r}{\left(
r^2- \left(1-\frac1{2K}\right) 
\right)} \, q(\theta) \cos(\theta),
\end{equation}
\end{lemma}
\medskip

\noindent
{\it Proof.} We have to minimize a quadratic functional under
the linear constraints $(1,u)=0$ and $u\in V_0$.   This corresponds to the
three constraints:
\begin{equation}
\label{eq:3c}
\int_0^{2\pi} u(\theta) \dd \theta \, =\,0,
\ \ \ \ \
\int_0^{2\pi} \cos(\theta) u(\theta) \dd \theta \, =\,0
\ \ \text{ and } \ \ 
\int_0^{2\pi} \sin(\theta)u(\theta) \dd \theta \, =\,0.
\end{equation}
The extrema (minima, by convexity) of such a problem 
can be found by the Lagrange multipliers method and they 
are of the form
\begin{equation}
\hat u (\theta) \, =\, -1 + \gl q(\theta) + \mu q(\theta) \cos(\theta)+
\eta   q(\theta)\sin (\theta),
\end{equation}
with $\gl$, $\mu$ and $\eta$ three real numbers.
The constraints \eqref{eq:3c}, via \eqref{eq:r} and \eqref{eq:sincos2},
yield the linear system $\eta=0$, $\gl+ \mu r=2\pi$ and $2K \gl r
+\mu (2K-1)=0$, which has a solution if and only if $2K(1-r^2)-1\neq 0$.
Since the minimum exists for every $K$ and since 
$2K(1-r^2)-1 \to 1$ for 
 $K\searrow 1$
we see that 
$2K(1-r^2)-1> 0$ for every $K$
(this is the upper bound in \eqref{eq:Bessel2}).
The proof is completed by making $\gl$ and $\mu$ explicit
and using that the expression in \eqref{eq:claim1} is
equal to $\gl^2+ 2 \gl \mu r+ \mu^2 (1-(1/2K))$.
\qed

\medskip

The following is one of our main statements: 
\medskip

\begin{proposition}
\label{th:stable}
 There exists $c_K \in (0,1/2)$ such that,
if $u \in L^2_{1/q}$ is such that $(1,u)= \int_0^{2\pi }u =0$, then
\begin{equation}
\label{eq:quasistable}
\cD(u)\, \ge \, c_K  \bra u-u_{1/2}, u-u_{1/2}\ket ,
\end{equation}
where $u_{1/2}:= q^\prime \bra u, q^\prime\ket/\bra q^\prime , q^\prime \ket$.
In particular, $\cD(u)\ge 0$.
\end{proposition}
\medskip

\noindent
{\it Proof.}
Lemma~\ref{th:decomp} shows that, if 
we write $u= v_0 + v_{1/2}+ \tilde v$
(according to the decomposition \eqref{eq:decomp}:
of course $v_{1/2}=u_{1/2}$) we have
\begin{equation}
\label{eq:Ddecomp}
\cD(u)\, =\, - (K-1) \bra \, \tilde v , \tilde v\, \ket + \frac 12 \bra v_0, v_0 \ket.
\end{equation}
We write $\tilde v (\theta)= \tilde c
q(\theta) \cos(\theta)$ and $v_0 =
a_0 +\sum_{j \ge 2} (a_j \cos(j\theta)+b_j \sin (j\theta))$
and, with this notations, one directly sees, using 
the definitions  \eqref{eq:q} and \eqref{eq:r}, that
the constraint $(1,u)=0$ is  equivalent to
\begin{equation}
\label{eq:explconstraint}
r \tilde c \, =\, -2 \pi a_0.
\end{equation}
Let us observe that we can assume $\tilde c \neq 0$: if $\tilde c=0$
then $v-v_{1/2}= v_0$ and, in view of \eqref{eq:Ddecomp},  
\eqref{eq:quasistable} holds. From now on we perform estimates
for arbitrary, but fixed, values of $\tilde c \neq 0$ (hence $a_0$ is fixed too).

Note then  that $\bra \tilde v, \tilde v \ket = \tilde c^2 (1-(1/(2K)))$,
by \eqref{eq:sincos2}, and we can therefore write 
\begin{equation}
\label{eq:Dstep1}
\cD(u)\, =\,
- (K-1)\left(1- \frac 1{2K}\right) \tilde c^2\, +
\, \frac{a_0^2}{2} \bra 1 +\frac{u_2}{a_0},  1 +\frac{u_2}{a_0} \ket,
\end{equation}
where of course we have set $u_2:= v_0 -a_0$.
By Lemma~\ref{th:prestable} we therefore obtain
that 
\begin{equation}
\label{eq:Dstep2}
\mintwo{u: (1,u)=0}{\text{ \ \ given }\tilde c } 
\cD(u)\, = \, \cD ( \tilde v + a_0 (1+ \hat u)) \, ,  
\end{equation}
where we recall that $\tilde c$ and $a_0$ are related via \eqref{eq:explconstraint}.  
Note that, by \eqref{eq:Bessel2},
$c_1(K)>0$ for $K>1$.
For sake of compactness let us introduce $\hat v_0:= a_0 (1+ \hat u) (\in V_0)$,
so that $u=(v_0-\hat v_0) + \hat v_0 +v_{1/2}+\tilde v$ 
and, looking back at \eqref{eq:Ddecomp}, we see that
\begin{equation}
\label{eq:Dstep3}
\begin{split}
\cD(u)\, &= \, - (K-1) \bra \, \tilde v , \tilde v\, \ket + \frac 12 \bra \hat v_0, \hat v_0 \ket
+  \bra \hat v_0, v_0-\hat v_0 \ket+ \frac 12 \bra v_0-\hat v_0, v_0-\hat v_0\ket
\\
&= \, \cD\left( \tilde v + \hat v_0 \right) +  \bra \hat v_0, v_0-\hat v_0 \ket+ \frac 12 \bra v_0-\hat v_0, v_0-\hat v_0\ket ,
\\
&= \, \mintwo{u: (1,u)=0}{\text{ \ \ given }\tilde c } 
\cD(u) +  \bra \hat v_0, v_0-\hat v_0 \ket+ \frac 12 \bra v_0-\hat v_0, v_0-\hat v_0\ket ,
\end{split}
\end{equation}
which implies in particular
\begin{equation}
\label{eq:non-neg}
 \bra \hat v_0, v_0-\hat v_0 \ket+ \frac 12 \bra v_0-\hat v_0, v_0-\hat v_0\ket \, \ge \, 0\, ,
\end{equation}
Since a lengthy computation yields
\begin{equation}
\frac{\cD\left( \tilde v + \hat v_0 \right) }
{
\bra \tilde v, \tilde v \ket +
\bra \hat v_0 , \hat v_0 \ket
}\, =\, 1 -K(1-r^2) \, =: c_K\, \in \, (0,1/2),
\end{equation}
where $c_K\in (0,1/2)$ is just a restatement
of \eqref{eq:Bessel2}, 
we get 
 \begin{equation}
 \begin{split}
 \cD(u) \, &=\,  c_K \left( \bra \tilde v , \tilde v\ket +
\bra \hat v_0 , \hat v_0 \ket \right)\, + \bra\hat  v_0, v_0-\hat v_0 \ket \, +\, 
\frac 12  
\, \bra v_0-\hat v_0, v_0-\hat v_0 \ket \, \\
&\ge c_K \left( \bra \tilde v , \tilde v\ket + 
\bra \hat v_0 , \hat v_0 \ket \right)\, +  2c_K \bra \hat v_0, v_0-\hat v_0 \ket \, +\, 
c_K
\, \bra v_0-\hat v_0, v_0-\hat v_0 \ket \\
& =\, c_K \left( \bra \tilde v , \tilde v\ket + 
\bra  v_0 ,  v_0 \ket \right),
\end{split}
\end{equation}
and the proof is complete.
\qed

\bigskip

For the next result we point out that,
by the definition of $H^{-1}_{1/q}$ in terms of rigged Hilbert spaces,
we know that there exists $c>0$ such that
 $\bbra u,u \kket \le c (u,u)$ for every
$u\in L^2$ with $(1,u)=0$. Therefore 
\begin{equation}
\label{eq:Poincare}
\bbra u,u \kket \le c_P^2 \bra u,u\ket\, ,
\end{equation}
with $c_P^2=c \max q$. This can be proven directly and $c_P$ can be made explicit,
in fact $\int \cU/q =0$ tells us that $\cU(\theta_0)=0$ for some $\theta_0$, so that 
$\bbra u,u \kket\le (\min q)^{-1} \int_{\theta_0}^{\theta_0+ 2\pi} \cU^2$
(we are  looking at $\cU$ as a periodic function with domain $\R$)
and the Poincar\'e inequality tells us that $\int_{\theta_0}^{\theta_0+ 2\pi} \cU^2$
is smaller than $\int_{\theta_0}^{\theta_0+ 2\pi} \left( \cU^\prime\right)^2=
(u,u)$. 
Therefore we can choose 
\begin{equation}
c_P^2\, 
= \, \frac{\max q}{\min q}\, =\, \exp(4Kr).
\end{equation}. 

\medskip

\begin{lemma}
\label{th:Francesco}
For $u \in L^2(\bbS)$ such that $(1,u)=0$ we have 
\begin{equation}
\label{eq:Francesco}
\Vert u - u_{1/2}\Vert_{L^2_{1/q}} \, \ge \, 
C \left\Vert u - \frac{\bbra u,q^\prime \kket}{\bbra q^\prime, q^\prime
\kket} q^\prime \right\Vert_{H_{-1,1/q}},
\end{equation}
where $C>0$ is given by
\begin{equation}
C^2\, :=\, \frac{ \left(1- (I_0(2Kr))^{-2}\right)}{2K r^2 c_P^4+ c_P^2 \left(1- (I_0(2Kr))^{-2}\right)}.
\end{equation}
\end{lemma}

\medskip

Of course Proposition~\ref{th:stable} and Lemma~\ref{th:Francesco}
yields the spectral gap inequality: for every $u$ such that
$\bbra u,q'\kket=0$ 
 we have
\begin{equation}
\label{eq:spectralgap}
\cD (u) \, \ge \, c_K C^2 \bbra u, u\kket.
\end{equation}

\medskip

\noindent{\it Proof.}
Set $e=q^\prime/\bra q^\prime, q^\prime \ket^{1/2}$, so that  $u_{1/2}=\bra u,e \ket e $.
By \eqref{eq:Poincare} we see that
\eqref{eq:Francesco} follows if one can show
\begin{equation}
\begin{split}
\Vert u - u_{1/2}\Vert_{L^2_{1/q}} \, &
\ge \, 
c_P C \left\Vert u - \frac{\bbra u, e \kket}{\bbra e,e
\kket} e \right\Vert_{L^2_{1/q}}\\
&= c_P C \sqrt{
\bra u-u_{1/2},u-u_{1/2}\ket + \left(
\bra u,e \ket - \frac{\bbra u,e \kket}{\bbra e,e \kket}
\right)^2
},
\end{split}
\end{equation}
 and this is equivalent (note that $c_P C \in (0,1)$) to
\begin{equation}
\label{eq:Francesco2}
\bra u-u_{1/2}, u-u_{1/2}\ket
\, \ge \, C_0 \left( \bra u, e\ket - \frac {\bbra u, e \kket}{\bbra e, e\kket}
\right)^2\, =\, 
C_0 \left(  \frac {\bbra u-u_{1/2}, e \kket}{\bbra e, e\kket}
\right)^2 ,
\end{equation}
with $C_0:= (c_PC)^2 /(1-(c_P C)^2)$.
By the Cauchy-Schwarz inequality and by \eqref{eq:Poincare} 
we have
\begin{equation}
\bbra u-u_{1/2},e \kket^2 \, \le\, 
 \bbra u-u_{1/2},u-u_{1/2}\kket \bbra e, e \kket \, \le\, c_P^2 
  \bra u-u_{1/2},u-u_{1/2}\ket \bbra e, e \kket,
  \end{equation}
  so that \eqref{eq:Francesco2}
  holds if $C_0 c_P^2 / \bbra e, e \kket \le 1$,
  which is equivalent to $c_P^2 C^2 \le \bbra e, e \kket /(c_P^2+
  \bbra e, e \kket)$. This is a condition on $C$ that we can verify explicitly by 
  using $ \bbra e, e \kket =(1- I_0^{-2}(2Kr))/(2Kr^2)$. 
  \qed
  
  \medskip

\subsection{Self-adjointness and spectral properties  of $L_q$}
Let us recall that $H$ is the space $L^2$ with zero average constraint
and that $L_q$ is viewed  as an operator on $H_{-1,1/q}\supset H$.
The first result is a technical lemma: 
 
 \begin{lemma}
 \label{th:1d-dir} Fix $K>1$. 
 There exists $c>0$ such that for every $u \in H$,
 we have
 \begin{equation}
 \bbra u, u\kket\, \ge \, c \, \Vert u_{1/2}\Vert_2^2, 
 \end{equation}
 where $u_{1/2}$ is the orthogonal projection, in $L^2_{1/q}$, on $V_{1/2}$.
 \end{lemma}
 
 \smallskip
 
 \noindent
 {\it Proof.}
 Using the explicit expression for $u_{1/2}$ (Proposition~\ref{th:stable}),
  and 
  \begin{equation}
  \bra  u , q^\prime \ket \, =\, \int (\log q )^\prime u \, 
  =\, - \int  (\log q )^{\prime\prime} \cU, 
  \end{equation}
  we see that
  \begin{multline}
  \Vert u_{1/2}\Vert_2^2 \, \le \, 
  (\max q) \bra u_{1/2} , u_{1/2} \ket \, =
  \\
   \frac{\max q }{\bra q^\prime, q^\prime \ket} \,  \left(
    \int  (\log q )^{\prime\prime} \cU \right)^2
    \, \le 
    \,
      \frac{2\pi \max q^2 \max \vert(\log q )^{\prime\prime}\vert^2 }{\bra q^\prime, q^\prime \ket} \,  \int   \cU ^2 /q \, =:\, \frac 1c \,\bbra u, u\kket ,
  \end{multline}
 where the last step is the definition of $c$.
 \qed
 
 \medskip
 
 \begin{proposition}
 \label{th:selfadj}
 $L_q$ is essentially self-adjoint.
 \end{proposition}

 \medskip
 
 \noindent
 {\it Proof.}
 We start by introducing for $u,v\in D(L_q)$
 \begin{equation}
 \label{cf:E1}
 \cE_1 (v,u) \, := \, \bbra v, (1-L_q)u \kket \, =\, 
 -\int v (\theta) \left(\int_0^\theta \frac{\cU}q\right)\dd \theta + \frac 12 \int \frac{v u} q - \int v \tilde J *u.
 \end{equation}
The right-most expression shows that $\cE_1(u,v)$ is well defined
as long as $u,v \in H$ (this generalizes the definition of $\cE_1(\cdot, \cdot)$ and we
will take this definition from now on). Moreover $\cE_1(\cdot, \cdot)$ is a continuous and coercive
bilinear form on $H\times H$, that is there exists
$c\in (0,1)$ such that
\begin{equation}
\label{eq:bandc}
\cE_1(u,v) \, \le \, \frac 1 c\Vert u \Vert_2 \Vert v \Vert_2, 
\ \ \text{ and } \ \ \ \cE_1(u,u) \, \ge \, c \Vert u \Vert_2^2.
\end{equation}
The second inequality follows from 
Proposition~\ref{th:stable} ad Lemma~\ref{th:1d-dir}. 
Now observe that for every $f \in H_{-1,1/q}$ the linear form 
$v \mapsto \bbra v, f \kket$, from $H$ to $\R$, is continuous
(it is continuous also as a map from $H_{-1,1/q}$ to $\R$) and therefore,
by the Lax-Milgram Theorem
\cite[Cor.~V.8]{cf:Brezis}, we  have that  
there exists a unique $u \in H$ such that 
\begin{equation}
\label{cf:LM}
\cE_1(v,u)\, =\,  \bbra v, f \kket, \ \ \text{ for every } v \in H. 
\end{equation}
Since 
we can write
\begin{equation}
\label{eq:cF1}
\bbra v, f \kket\, =\, - \int v (\theta) \left( \int_0^\theta \frac{\cF}q \right)\dd \theta,
\end{equation}
from \eqref{cf:E1}, \eqref{eq:cF1} and \eqref{cf:LM}
we see that
\begin{equation}
\label{eq:tbr}
-\int_0^\theta \frac{\cU}q + \frac{u(\theta)}{2q(\theta)} -
\left( \tilde J * u \right) (\theta) \, =\, -  \int_0^\theta \frac{\cF}q,
\end{equation}
for (Lebesgue) almost every $\theta$. Since $u\in H$, the primitive
of $\cU /q$ is $C^1$ and its (weak) second derivative is square integrable.
If $f\in H (\subset H_{-1, 1/q})$ then the same is true for the right-hand side in 
\eqref{eq:tbr}. Since $ \tilde J * u$ is $C^\infty$, we see that $u$ is $C^1$
(more precisely, has a $C^1$ version). So the left-most term in \eqref{eq:tbr}
is at least $C^3$. If now we assume that $f$ is $C^0$, we can therefore 
conclude that $u\in C^2$. 

To sum up: if $f$ is periodic, $C^0$, with $\int f =0$, then $u \in D(L_q)$ and
$(1-L_q) u = f$, which follows by taking applying
$\partial_\theta (q(\theta) \partial_\theta\,  \cdot\,  )$ to both terms in  \eqref{eq:tbr} (and by using
\eqref{eq:convstat}). Since such functions $f$  are dense in $H_{-1,1/q}$,
we see that the range of $1-L_q$ is dense, so that its kernel is $\{0\}$, 
and this implies that
$L_q$ is essentially self-adjoint (\cite[Prop.~VII.6]{cf:Brezis}).
\qed 

\medskip

\begin{proposition}
\label{th:purepoint}
The spectrum of $L_q$ is pure point.
\end{proposition}
\medskip

\noindent
{\it Proof.}
We are going to prove this by showing that 
the resolvent of $L_q$ is compact, namely that 
$(\gl -L_q)^{-1}$ is compact for $\gl$ in the resolvent set. 
It suffices to prove such a result for one value of
$\gl$ \cite[p.~187]{cf:Kato} and we choose $\gl=1$, which is in the resolvent set
thanks to Proposition~\ref{th:stable} and Proposition~\ref{th:selfadj}.
So let us consider $f:=(1 -L_q)^{-1} u$, $u \in H_{-1,1/q}$,
so that $f$ is in the domain of $1-L_q$ and we have
\begin{equation}
\bbra f , (1-L_q) f \kket \, = \, \bbra f, u \kket.
\end{equation}
But, by \eqref{eq:bandc}, $\bbra f , (1-L_q) f \kket$ is bounded below by 
$c \Vert f \Vert _2^2$, so that
\begin{equation}
c \, \Vert f \Vert _2 \, \le \, \frac{\bbra f, u \kket}{\Vert f \Vert _2} \, \le \,
\frac 1C
\frac{\bbra f, u \kket}{\sqrt{\bbra f, f \kket }} \, \le \frac 1C \sqrt{\bbra u, u\kket},
\end{equation}
where we have used the continuous injection of $H$ into $V'=H_{-1,1/q}$
($C$ is the constant arising when comparing the norms of these two spaces).
Therefore $(1-L_q)^{-1}$ maps sequences that are bounded in
$H_{-1,1/q}$ to sequences that are bounded in $H$. We
are therefore left with showing that the embedding of $H$ into
$H_{-1,1/q}$ is compact.
This just follows by the Cauchy-Schwarz inequality: for every 
$v\in H$ we have
$\vert \cV (\theta )- \cV (\theta^\prime) \vert \le \Vert v \Vert _2
\sqrt{\vert \theta - \theta^\prime\vert} $ and , since we are on a bounded
interval with periodic boundary conditions and $\int \cV/q =0$, this
yields that  $\{ \cV: \, v \in H $ and $\Vert v \Vert _2\le const.\}$
is a compact subset of $C^0$ (Ascoli-Arzel\'a Theorem), 
and hence of $L^2_{1/q}$.
That is,  a bounded subset of $H$ is a relatively compact subset of 
$H_{-1,1/q}$ and 
this completes the proof.
\qed

\section{The nonlinear evolution}
\label{sec:nonlinear}

 
We need the following result on  the nonlinear evolution:

\medskip

\begin{proposition}
\label{th:nonlinear}
For every  $\nu_0 \in \cM_1(\bbS)$
 there is a unique element 
$\nu_\cdot $ of $ C^0 ([0,\infty); \cM_1(\bbS))$ such that
\eqref{eq:limitweak} is satisfied for every $F \in C^2(\bbS)$ and 
 every $t>0$. Moreover
for $t>0$  the measure $\nu_t$ is absolutely continuous
with respect to the Lebesgue measure and,  
if we denote by $q_t(\cdot)$  its density, the function $
(t, \theta)\mapsto q_t(\theta)$, with domain of definition
$(0, \infty) \times \bbS$, is $C^\infty$,  strictly positive and
it solves \eqref{eq:classical}.
\end{proposition}

\medskip

\noindent
{\it Proof.}
The uniqueness result is proven for example in \cite{cf:Oelsch}
(with $\bbS$ replaced by $\R$) and in \cite{cf:Gartner} (for a general,
bounded or unbounded,
domain subset of $\R^d$). The case of periodic boundary conditions
is not treated explicitly but the argument of proof goes through 
with minor modifications. Existence is established by the tightness 
result on the particle system, which follows
by methods that are by now standard: it suffices in fact
to show that, for every (smoooth) $F$, $\{ \int_\bbS F \dd \nu_{N,t}\}_N$
is tight 
(see {\sl e.g.} \cite{cf:KL,cf:Oelsch,cf:Gartner})
and this follows immediately from the fact that the drift in \eqref{eq:Kuramoto0}
is bounded.

For the regularity and positivity aspects we use the fact that,
if for $t>0$ there is a solution to
\begin{equation}
\label{eq:semiclassical}
\partial_t u(t,\theta) \, =\, \frac 12 \frac{ \partial ^2 u (t,\theta)}{\partial \theta^2}
+ K  \frac{\partial}{\partial \theta}\left[\left(
H(t,\theta)
\right)
u(t,\theta)\right] ,
\end{equation}
with 
\begin{equation}
H(t,\theta) \, :=\, \int_{\bbS} \sin(\theta- \theta^\prime ) \nu_t(\dd \theta^\prime)
\, =\, \sin (\theta)  \int_{\bbS} \cos(\theta^\prime ) \nu_t(\dd \theta^\prime)
-\cos (\theta)  \int_{\bbS} \sin(\theta^\prime ) \nu_t(\dd \theta^\prime)
, 
\end{equation}
(note that $H(\cdot, \cdot)$ is continuous in time and $C^\infty$ in space)
such that for every $F \in C^0 (\bbS)$
\begin{equation}
\lim_{t\searrow 0} \int_\bbS F(\theta) u (t, \theta) \dd \theta
\, =\, \int F \dd \nu_0,
\end{equation}
then one directly verifies that $\tilde  \nu _\cdot \in C^0([0,T];
\cM_1(\bbS))$,
defined by $\tilde \nu_t(\dd\theta) =u(t, \theta) \dd\theta$
for $t>0$
and $\tilde \nu_0=\nu_0$, is a solution 
to \eqref{eq:limitweak}. Hence $\tilde \nu_\cdot =\nu_\cdot$ by uniqueness.

Equation \eqref{eq:semiclassical} is a parabolic linear partial differential equation
on which there is much literature. For our purpose the results by D. G. Aronson in \cite{cf:Aronson}
 turn out to
be particularly relevant. In particular, Aronson
shows that the equation \eqref{eq:semiclassical} (consider 
$\theta \in \R $ for the moment)
admits a fundamental solution
$\Gamma (t, \theta; s, \theta^\prime)$ ($t>s$)
so that the solution at time $t$ can be written as 
$\int_\R \Gamma (t, \theta; 0, \theta^\prime) u(0, \theta^\prime) \dd \theta^\prime$,
at least when $u(0, \cdot)\in L^2_{\text{loc}}(\bbR)$
(of course the solution has to be interpreted in
a weak sense: see  \cite[pp. 608-609]{cf:Aronson}
for the very general set up in which such a result it is proven).
A number of results on the fundamental solution are proven
\cite[Section~7]{cf:Aronson} and notably that it is a continuous function
in $(t, \theta)$ for $t>s$ and that,
if we assume that 
$\sup_{t\in[s,s+T], \theta \in \R}H(t, \theta)=:M< \infty$,  there exists $C=C(T,M)>0$
such that
\begin{equation}
\label{eq:Aronson}
\frac 1{C\sqrt{t-s}}
\exp\left(-  \frac{C (\theta -\theta^\prime)^2}{\sqrt{t-s}}\right)\, \le \, 
\Gamma (t, \theta; s, \theta^\prime)\, \le \, 
\frac C{\sqrt{t-s}} \exp\left(-  \frac{(\theta -\theta^\prime)^2}{C\sqrt{t-s}}\right),
\end{equation}
for every $\theta$, $\theta^\prime$ and every $t\in (s, s+T]$.
Moreover (\cite[Corollary~12.1]{cf:Aronson}) the result applies 
also to initial data that are measures: namely, if $\mu$ is a measure on 
$\R$ such that  for $t>0$ the map $\theta \mapsto
\int_\R \Gamma (t, \theta; 0, \theta^\prime) \mu (\dd \theta)$
is a function in $L^2_{\text{loc}}(\bbR)$ then such a map defines
a weak solution to \eqref{eq:semiclassical} on $(0, T]\times \R$
(namely a weak solution to \eqref{eq:semiclassical} on
$[t_0, T]\times \R$ for every $t_0 \in (0, T)$).

Let us now specialize to our case: $H(t, \cdot)$ is smooth and $2\pi$-periodic,
so that 
$\Gamma (t, \theta; s, \theta^\prime)= \Gamma (t, \theta+2\pi; s, \theta^\prime+2\pi)$.
Let us apply the result we have just stated with $\mu$ defined
by requiring that its restriction to $[2\pi j , 2\pi (j+1))$
coincides with the image of the measure $\nu_0$ under the 
application $[0, 2\pi)\ni \theta \mapsto \theta+ 2\pi j \in [2\pi j , 2\pi (j+1))$,
for every $j$.
Therefore for $t>0$
\begin{equation}
\label{eq:Aronson-per}
\int_\R \Gamma (t, \theta; 0, x) \mu (\dd x)\, =\, 
\sum_{j \in \bbZ} \int_{[0, 2\pi)} \Gamma (t, \theta; 0, \theta^\prime+2\pi j) 
\nu (\dd \theta^\prime)\, =:\, v(t, \theta),
\end{equation}
and, by 
\eqref{eq:Aronson}, $v(t, \cdot)$
is bounded  as soon as $t>0$.
Therefore $v(\cdot, \cdot)$ is a weak solution and, in turns,
$v(t, \theta) \dd \theta$ coincides with $\nu_t(\dd \theta)$,
which therefore has a representation in terms
of the fundamental solution $\Gamma$
and this implies not only that the solution becomes a 
bounded function as soon as $t>0$, but also (by the lower bound
in \eqref{eq:Aronson}) that it is strictly positive.

At this point, since we know that the solution 
is bounded,  the smoothness in both variables of the solution
(for $t>0$) may be derived by standard methods:
this issue is taken up for example in \cite{cf:AG}
for a slightly different evolution equation, or, more generally, in \cite{cf:Robinson}. 
\qed

\section{On the irreversibility of the Kuramoto model}
\label{sec:irreversibility}

In this section we consider the Kuramoto $h$-model, {\sl i.e.} \eqref{eq:Kuramoto}
with {\sl general} drift, as in Remark
\ref{rem:h}, and $\gs =1$. 

Existence of a unique invariant probability measure (for each fixed $N$)
is a well known fact, but one can actually prove that
such an invariant measure has a  positive $C^\infty$ density $\gr: \bbS^N \to (0,\infty)$.
These issues are treated in detail for example in \cite[Ch.s~3 and 5]{cf:JQQ},
where one finds also an extensive treatment of the entropy productions
for Markov processes (with references to the vast literature on the subject).
The entropy production rate $e_p$ for a  stationary process $X$  is defined 
as the limit as $T \to \infty$, when it exists, of the relative entropy of
the law of $\{X_t\}_{t\in [0, T]}$ with respect to the law of 
$\{X_{T-t}\}_{t\in [0, T]}$, divided by $T$. 
For a large class of models $e_p$ takes the form of the steady state average of 
time integral of the square of a suitable {\sl flux}. This is true also in our case,
namely 
 $e_p=
\frac 12 \int_{\bbS^N} \sum_{j=1}^N
J_j\left(\underline{\gp}\right)^2 \rho ( \underline{\gp}) \dd \underline{\gp}$, where 
\begin{equation}
\label{eq:ent-flux}
J_j\left(\underline{\gp}\right)\, :=\, 2\xi_j -
\frac 2N \sum_{i=1}^N h( \gp_j -\gp_i) - \frac{\partial}{\partial \gp_j}  \log\rho\left(\underline{\gp}\right). 
\end{equation}
The key point is that $e_p=0$
if and only if the system is reversible
\cite[Th.~5.4.6]{cf:JQQ} (of course reversibility 
 calls for  specifying an  invariant probability  with respect to which
the system is reversible, but  in our set-up there is only one
invariant measure). Therefore our system is reversible
 if and only if $J_j(\cdot) \equiv 0$ for every
$j$: let us spell it out
\begin{equation}
\label{eq:rev1}
  \frac{\partial}{\partial \gp_j}  \log\rho\left(\underline{\gp}\right)\, =\, 2\xi_j -
\frac 2N \sum_{i=1}^N h( \gp_j -\gp_i)\ \ \text{ for every } j \text{ and } \underline{\gp}.
\end{equation}
This expression directly implies that $\int_\bbS h(\theta)\dd \theta =2\pi \xi_j$
for every $j$, that is  $\xi_j$ does not depend on $j$.
Without loss of generality we may therefore assume $\xi_j=0$ 
for every $j$ (recall Remark~\ref{eq:wirr}), which entails
$\int_\bbS h(\theta)\dd \theta=0$
and  therefore the primitive $\tilde h$ of $h$ is $2\pi$-periodic (we make the
arbitrary choice $\tilde h(0)=0$).
By integrating \eqref{eq:rev1}
we obtain that
\begin{equation}
\label{eq:rev2}
\log \rho (\gp) \, =\, \frac 2N \sum_i \tilde h(\gp_j- \gp _i) +c_j \left(\underline{\gp}\right),
\end{equation}
where $c_j \left(\underline{\gp}\right)$ does not depend on $\gp_j$.
For  $j=1$ and $\gp_i$ fixed for $i=3,4, \ldots$
we can rewrite \eqref{eq:rev2}
as 
\begin{equation}
\label{eq:rev3}
\log \rho (\gp_1, \gp_2) \, =\, \frac 2N \tilde h(\gp_1- \gp _2)+
g_1(\gp_1) +g_2 (\gp_2)\, ,
\end{equation}
where $g_1(\gp_1):=(2/N) \sum_{i\ge 3} \tilde h(\gp_j- \gp _i)$ and 
$g_2(\gp_2):=c_1 \left(\underline{\gp}\right)$.
We can of course repeat the same steps with
$j=2$ obtaining thus 
\begin{equation}
\label{eq:rev4}
\log \rho (\gp_1, \gp_2) \, =\, \frac 2N \tilde h(\gp_2- \gp _1)+
f_1(\gp_1) +f_2 (\gp_2)\, ,
\end{equation}
with $f_1$ and $f_2$ defined in analogy with $g_1$, $g_2$
(but we are simply interested in the fact that they are smooth functions from
$\bbS$ to $\bbR$). From \eqref{eq:rev3} and \eqref{eq:rev4} we infer that
\begin{equation}
\tilde h(\gp_1- \gp _2)- \tilde h(\gp_2- \gp _1)\, =\, f(\gp_1)+ g(\gp_2),
\end{equation}
for suitable smooth functions $f$ and $g$ from $\bbS$ to $\bbR$.
This tells us in particular that $f(c+\theta)+g(c)$ does not depend
on $c$ (it is equal to $\tilde h (\theta)-\tilde h (-\theta)$)
 and, therefore, that $f(x)+g(x)$ is a constant and $f(c+x)-f(x)=f(c)-f(0)$
for every $c$ and $x$ (that is $f$ is constant, since it is continuous
and periodic). We have therefore reached the conclusion
that $\theta \mapsto \tilde h(\theta)- \tilde h(-\theta)$ is a constant,
which has therefore to be zero. 

We sum up the argument we have just developed in the
following statement:

\medskip

\begin{proposition}
For every $N$
the dynamics defined by \eqref{eq:Kuramoto}, generalized as in 
Remark~\ref{rem:h}, is reversible if and only if the following two conditions are satisfied:
\begin{enumerate}
\item $\xi_1=\xi_j$ for every $j$;
\item $h(\cdot)-\xi_1: \bbR\mapsto \bbR$ is an odd function.
\end{enumerate}
\end{proposition}

\section*{Acknowledgments}
We are very grateful to Francesco Caravenna for suggesting
Lemma~\ref{th:Francesco}. G.G. thanks also Marek Biskup, Fran\c{c}ois
Delarue and Lorenzo Zambotti for fruitful suggestions and discussions. This
work has been supported by the ANR grant MANDy.

\end{document}